\documentclass[aps,prc,twocolumn,preprintnumbers,amsmath,amssymb,floatfix]{revtex4-2}

\usepackage{graphicx}
\usepackage{bm}
\usepackage{hyperref}

\begin{document}

\title{Inclusive breakup reactions with nonspectator fragments: Generalization of the Ichimura-Austern-Vincent sum rules}

\author{Jin Lei}
\email[]{jinl@tongji.edu.cn}
\affiliation{School of Physics Science and Engineering, Tongji University, Shanghai 200092, China.}
\affiliation{Southern Center for Nuclear-Science Theory (SCNT), Institute of Modern Physics, Chinese Academy of Sciences, Huizhou 516000, Guangdong Province, China.}

\date{\today}

\begin{abstract}
The Ichimura-Austern-Vincent (IAV) sum rule formalism for inclusive
breakup reactions $a + A \to b + \mathrm{anything}$ treats the
detected fragment $b$ as a spectator by replacing its interaction
with the target by an optical potential.
This assumption becomes questionable when $b$ is a loosely bound
composite particle such as a deuteron.
I derive a generalization that removes the spectator approximation
and retains $b$'s internal degrees of freedom, providing
state-resolved inclusive cross sections.
Within the DWBA, all nonspectator effects enter through the source
function via the operator $V_{bA} - U_{bA}$.
The exact sum rule involves the full $x + A$ resolvent
$(E_{x,0}^+ - H_{xA})^{-1}$, while a single-channel IAV-like
expression is recovered only when the explicit target dependence of
$V_{bA}$ is neglected; post-prior equivalence is preserved in both
cases.
A key conceptual finding is that the standard IAV result for
structureless $b$ corresponds, under closure and a reduction of
$b$'s internal space, to
the \emph{total} inclusive cross section summed over all of $b$'s
internal states, rather than the cross section for $b$ in a
specific state.
An operator-level estimate for $b = d$ on ${}^{208}\mathrm{Pb}$
shows that the nonspectator correction is not a small perturbation
at the nuclear surface.
The present work is purely formal: it establishes the theoretical
framework and identifies the relevant operators, while quantitative
assessment of the cross-section impact awaits a full numerical
evaluation of the source integrals.
\end{abstract}

\maketitle

\section{Introduction}
\label{sec:intro}

Inclusive breakup reactions, in which a composite projectile $a$ impinges
on a target $A$ and only one fragment $b$ is detected in the exit channel
while everything else is summed over, play an important role in nuclear
reaction theory~\cite{Moro2025}.
The reaction $a + A \to b + \mathrm{anything}$, where $a = b + x$,
encompasses a range of physical processes: elastic breakup (EBU), in which
both fragments emerge with the target in its ground state;
nonelastic breakup (NEB), also called breakup fusion, in which fragment $x$
is absorbed by the target; and more complex rearrangement
channels~\cite{Pampus1978,Wu1979,Matsuoka1980}.
Because only fragment $b$ is detected, the experiment integrates over all
of these final-state configurations, making a fully exclusive theoretical
description unnecessary and motivating sum-rule approaches that exploit
this inclusiveness.

The theoretical framework for inclusive breakup was developed in a series
of papers by Udagawa, Tamura, and collaborators~\cite{Udagawa1981,Li1984}
(the prior-form or UT approach), and independently by Austern and
Vincent~\cite{Austern1981} and Ichimura, Austern, and
Vincent~\cite{Ichimura1985} (the post-form or AV approach).
These two lines of work, now collectively referred to as the IAV formalism,
derive closed-form sum rules for the inclusive breakup cross section
by performing the sum over unobserved final states analytically through
quantum mechanical completeness.
The central result of the post-form derivation is that the inclusive
nonelastic breakup cross section can be expressed in terms of the imaginary
part of an optical-model Green's function $G_x$ for the unobserved
fragment $x$ in the target field, acting on a source function that encodes
the breakup dynamics.
Ichimura, Austern, and Vincent demonstrated that the post and prior forms
of the sum rule are equivalent, provided certain surface terms vanish
upon appropriate regularization~\cite{Ichimura1985,Kasano1982}.

A key assumption underlying the IAV formalism is that the detected
fragment $b$ acts as a \emph{spectator}.
In the model Hamiltonian of Ref.~\cite{Ichimura1985} [their Eq.~(2.1)],
the interaction $V_{bA}$ between $b$ and the target is replaced by an
optical potential $U_{bA}$ that depends only on the center-of-mass
coordinate of $b$ and does not excite the target.
The distorted wave $\chi_b^{(-)}$ for $b$ in the exit channel is
generated by a (generally different) optical potential $U_{bB}$.
The post-form residual interaction in the IAV model is then
$V_{\mathrm{post}}^{\mathrm{IAV}} = V_{bx} + U_{bA} - U_{bB}$, which
does not depend on the target internal coordinates since neither
$U_{bA}$ nor $U_{bB}$ excite the target.
This property is essential for the Feshbach optical
reduction~\cite{Feshbach1958,Feshbach1962}
that produces the closed-form sum rule involving $G_x$.
In addition, $b$ is treated as structureless, eliminating any role
for its internal degrees of freedom.
For reactions where $b$ is indeed a tightly bound, compact fragment
(such as an alpha particle), the spectator approximation is well justified,
and the IAV formalism has been applied with considerable
success~\cite{Potel2015,Potel2017,Lei2015,Lei2015b,Lei2017,Carlson2016,Lei2019,Lei2019b,Lei2016review}.

There exist, however, physically important reactions in which the detected
fragment $b$ is itself a loosely bound composite
particle~\cite{Canto2006,Keeley2007}.
A prominent example is inclusive deuteron production in reactions induced
by ${}^6\mathrm{Li}$, ${}^7\mathrm{Li}$, or triton projectiles, where
$b = d$ (deuteron) and $x$ is a heavier fragment absorbed by the target.
The deuteron, with a binding energy of only 2.224~MeV and a root-mean-square
proton-neutron separation of approximately 3.9~fm, is highly susceptible
to polarization, excitation, and breakup in the nuclear and Coulomb fields
of the target~\cite{Johnson1970,Rawitscher1974,Austern1987}.
An even more extreme case arises in ${}^9\mathrm{Be}$-induced
reactions~\cite{Villanueva2024}, where the detected fragment
$b = {}^8\mathrm{Be}$ ($\alpha + \alpha$) is itself unbound and
necessarily breaks up after neutron transfer to the target.
In such cases, the replacement $V_{bA} \to U_{bA}$ is a significant
approximation, because the full interaction $V_{bA}$ depends on the
internal coordinates of $b$ and can couple $b$'s internal excitations
to target excitations.
Ichimura, Austern, and Vincent themselves noted this limitation, remarking
in Sec.~V of Ref.~\cite{Ichimura1985} that
``the importance of excitation during breakup should be no surprise''
and that ``a reduced theory of a many-body system, in terms of the
coordinates of only a few `active' particles, requires the careful
introduction of effective interactions.''
Indeed, an effective three-body Hamiltonian analysis based on double
Feshbach projection~\cite{Lei2026} shows that the standard additive
model for composite projectiles misses non-additive induced
interactions whose omission can significantly distort reaction
cross sections.
Despite this recognition, no systematic generalization of the IAV sum rules
to the nonspectator case has been developed.

The practical implementation of the IAV formalism has advanced significantly
in recent years.
Potel, Nunes, and Thompson~\cite{Potel2015,Potel2017} implemented the
prior-form sum rule and demonstrated that it provides a convergent,
practical scheme for computing inclusive breakup cross sections.
Lei and Moro~\cite{Lei2015,Lei2015b,Lei2017} extended these calculations
to a wide range of reactions and showed good agreement with experimental
data for $(d,p)$-type and $({}^6\mathrm{Li}, \alpha)$-type reactions,
and further demonstrated that the IAV framework can explain complete
fusion suppression and partial fusion in weakly bound
systems~\cite{Lei2019,Lei2019b}.
These implementations exclusively use the prior form, because the post-form
matrix element suffers from a convergence issue related to the disconnected
part of the elastic breakup amplitude, as discussed in detail in
Refs.~\cite{Ichimura1985,Lei2025post}.
The prior form avoids this problem and provides numerically stable
results.
Carlson, Frederico, and Hussein~\cite{Carlson2017} extended the
spectator model to three-fragment projectiles within a four-body
framework, obtaining inclusive elastic and nonelastic breakup cross
sections for weakly bound systems such as Borromean nuclei.
More recently, Neoh \textit{et al.}~\cite{Neoh2016} formulated inclusive
breakup within the eikonal reaction theory, and
Deltuva~\cite{Deltuva2025} developed a Faddeev-based
inclusive breakup framework that treats all three-body channels on
equal footing.

In this paper, I derive a generalization of the IAV inclusive breakup
formalism that removes the spectator approximation for fragment $b$.
A key conceptual point, developed in Sec.~\ref{sec:relation}, is that
the standard IAV treatment with structureless $b$ admits a natural
correspondence, under closure and structureless-reduction
approximations, with
the \emph{total} inclusive cross section summed over all of $b$'s
internal states, rather than the
state-resolved cross section for detecting $b$ in a specific state.
The present formalism introduces the projection onto $b$'s ground
state explicitly, providing access to state-resolved observables.
At the exact DWBA level the resulting sum rule involves the full
$x + A$ resolvent, whereas the familiar optical-propagator form
reappears only in a reduced single-channel limit.
Following the logical structure of the original IAV paper, I first
develop the post-form sum rule in Sec.~\ref{sec:post}, then the
prior-form sum rule and post-prior equivalence in
Sec.~\ref{sec:prior}.
In both forms, the nonspectator correction is controlled by the
single operator $V_{bA} - U_{bA}$.
In Sec.~\ref{sec:estimate}, I present a tidal estimate for the
deuteron case.
I discuss practical implications in Sec.~\ref{sec:discussion} and
summarize in Sec.~\ref{sec:summary}.

\section{Post-form sum rule}
\label{sec:post}

I consider the inclusive breakup reaction $a + A \to b + \mathrm{anything}$,
where the projectile $a$ is a bound state of fragments $b$ and $x$,
so that $a = b + x$.
Fragment $b$ is composite with internal degrees of freedom $\zeta$,
described by the internal Hamiltonian $h_b(\zeta)$ with eigenstates
$h_b |\phi_n\rangle = \varepsilon_n |\phi_n\rangle$.
The ground state $|\phi_0\rangle$ with energy $\varepsilon_0$ is the
state in which $b$ is detected experimentally, while the states with
$n \geq 1$ include excited bound states and continuum states
corresponding to the breakup of $b$.
These eigenstates form a complete set,
$\sum_n |\phi_n\rangle\langle\phi_n| = \mathbf{1}_\zeta$,
in $b$'s internal Hilbert space.

The full Hamiltonian for the $a + A$ system is
\begin{equation}
H = H_A(\xi) + h_b(\zeta) + K_b + K_x + V_{bx} + V_{bA} + V_{xA} \,,
\label{eq:H_full}
\end{equation}
where $H_A(\xi)$ is the internal Hamiltonian of the target nucleus with
ground state $H_A \Phi_A = E_A \Phi_A$, $\xi$ denotes the target internal
coordinates, $K_b$ and $K_x$ are the kinetic energies defined below,
and $V_{bx}(\mathbf{r}_{bx}, \zeta)$,
$V_{bA}(\mathbf{r}_{bA}, \zeta, \xi)$, and $V_{xA}(\mathbf{r}_x, \xi)$
are the pairwise interactions.
The Hamiltonian is written in the center-of-mass frame using relative
coordinates, retaining the full target mass with no heavy-target
approximation.
I use the position $\mathbf{r}_x$ of $x$ relative to the target $A$,
the position $\mathbf{r}_{bA}$ of $b$ relative to $A$, and the position
$\mathbf{r}_{bB}$ of $b$ relative to the center of mass of the residual
$B = x + A$.
The $b$-$x$ relative coordinate is
$\mathbf{r}_{bx} = \mathbf{r}_{bA} - \mathbf{r}_x$, and the projectile
center of mass relative to $A$ is
$\mathbf{r}_a = (m_b \mathbf{r}_{bA} + m_x \mathbf{r}_x)/m_a$, with
$m_a = m_b + m_x$.
The two $b$ coordinates differ by the recoil of the residual,
$\mathbf{r}_{bA} = \mathbf{r}_{bB}
+ \tfrac{m_x}{m_x + M_A}\,\mathbf{r}_x$, and coincide as
$M_A \to \infty$.
Each binary interaction is local in its respective separation:
$V_{bA}$ and $U_{bA}$ in $\mathbf{r}_{bA}$ (with $V_{bA}$ additionally
parametric in $\zeta$ and $\xi$, as detailed below), $V_{xA}$ in
$\mathbf{r}_x$, $V_{bx}$ in $\mathbf{r}_{bx}$ (and in $\zeta$), and the
auxiliary distorting potential $U_{bB}$ (defined below) in
$\mathbf{r}_{bB}$.
The kinetic energies form two equivalent Jacobi pairs: in the exit
channel, $K_x$ for $x$ relative to $A$ and $K_b$ for $b$ relative to
$B$, conjugate to $\mathbf{r}_x$ and $\mathbf{r}_{bB}$; in the entrance
channel, $K_{bx}$ for the $b$-$x$ motion and $K_a$ for the projectile
$a = b + x$ relative to $A$, conjugate to $\mathbf{r}_{bx}$ and
$\mathbf{r}_a$.
Each pair diagonalizes the kinetic energy with no cross term, and the
two are connected by the exact Jacobi transformation
$K_b + K_x = K_{bx} + K_a$ used in the prior form (Sec.~\ref{sec:prior}).
To keep the expressions compact, the arguments of operators and wave
functions are displayed only at their first appearance or where the
explicit dependence is essential to the argument, and are suppressed
otherwise.
The interaction $V_{bx}$ depends on $\mathbf{r}_{bx}$ and on $\zeta$
because the binding between $b$ and $x$ is sensitive to $b$'s
internal configuration.
The interaction $V_{bA}$ depends on $b$'s center-of-mass position
$\mathbf{r}_{bA}$, on $b$'s internal coordinates $\zeta$ (because the
individual constituents of $b$ interact with the target at different
positions), and on $\xi$ (because it can excite the target).
The projectile bound-state wave function
$\varphi_a(\mathbf{r}_{bx}, \zeta)$ satisfies
\begin{equation}
(h_b + K_{bx} + V_{bx})\varphi_a = -\epsilon_a \varphi_a \,,
\label{eq:bound_state}
\end{equation}
where $K_{bx}$ is the $b$-$x$ relative kinetic energy and $\epsilon_a > 0$
is the projectile binding energy.

For comparison, the IAV model Hamiltonian~\cite{Ichimura1985} is
$H^{\mathrm{IAV}} = H_A + K_b + K_x + V_{xA} + U_{bA} + V_{bx}$,
which differs from Eq.~(\ref{eq:H_full}) in that $V_{bA}$ has been
replaced by an optical potential $U_{bA}$ and $h_b$ is absent.

To construct the post-form DWBA, I introduce the exit-channel Hamiltonian
\begin{equation}
H_{\mathrm{exit}} = h_b + K_b + U_{bB} + H_{xA} \,,
\label{eq:H_exit}
\end{equation}
where $H_{xA} = H_A + K_x + V_{xA}$ is the full $x + A$ Hamiltonian
and $U_{bB}$ is an auxiliary optical potential that generates the distorted
wave $\chi_b^{(-)}$ for fragment $b$.
The exit-channel Hamiltonian is separable in $b$'s degrees of freedom
($h_b + K_b + U_{bB}$) and the $x + A$ system ($H_{xA}$), so the
exit-channel eigenstates factorize as
$|\chi_b^{(-)}(\mathbf{k}_b)\rangle\,|\phi_n(\zeta)\rangle\,
|\Psi_{xA}^c\rangle$,
where $H_{xA}|\Psi_{xA}^c\rangle = E^c |\Psi_{xA}^c\rangle$.
The post-form residual interaction is
\begin{equation}
V_{\mathrm{post}} = H - H_{\mathrm{exit}} = V_{bx} + V_{bA} - U_{bB} \,.
\label{eq:Vpost}
\end{equation}
In the IAV model, where $V_{bA}$ is replaced by $U_{bA}$,
the post-form residual is
$V_{\mathrm{post}}^{\mathrm{IAV}} = V_{bx} + U_{bA} - U_{bB}$,
which does not depend on the target coordinates $\xi$ since
neither $U_{bA}$ nor $U_{bB}$ excite the target.
In the present case, $V_{\mathrm{post}}$ contains the full
interaction $V_{bA}$, which depends on $\mathbf{r}_{bA}$, $\zeta$,
and $\xi$.

The DWBA transition amplitude for detecting $b$ in its ground state
$\phi_0$ while the $x + A$ system is in eigenstate $c$ is
\begin{equation}
T_{0,c} = \langle \chi_b^{(-)} \phi_0\, \Psi_{xA}^c |\,
V_{\mathrm{post}} \,| \chi_a^{(+)} \varphi_a \Phi_A \rangle \,,
\label{eq:T0c}
\end{equation}
where $\chi_a^{(+)}$ is the entrance-channel distorted wave.
The doubly differential inclusive cross section, summed over all
unobserved final states of the $x + A$ system, is
\begin{equation}
\frac{d^2\sigma}{dE_b\, d\Omega_b}\bigg|_{\mathrm{post}}
= \frac{(2\pi)^4}{v_a}\,
\sum_c |T_{0,c}|^2\, \delta(E_{x,0} - E^c) \,,
\label{eq:xsec_sum}
\end{equation}
where $v_a$ is the $a$-$A$ relative velocity,
$E^c$ is the eigenenergy of the $x + A$ final state $\Psi_{xA}^c$,
and $E_{x,0} = E - E_b - \varepsilon_0$ is the energy available
to the $x + A$ system when $b$ is in its ground state with
kinetic energy $E_b = \hbar^2 k_b^2/(2\mu_b)$, with $\mu_b$ the
$b$-$(x\!+\!A)$ reduced mass.

I define the source function by integrating out $b$'s degrees of freedom,
\begin{equation}
|\rho_0\rangle = ( \phi_0(\zeta)\, \chi_b^{(-)}(\mathbf{r}_{bB})|\,
V_{\mathrm{post}} \,| \chi_a^{(+)} \varphi_a \Phi_A \rangle \,,
\label{eq:source}
\end{equation}
which is obtained by projecting $V_{\mathrm{post}}|\chi_a^{(+)}\varphi_a
\Phi_A\rangle$ onto $b$'s ground state and the distorted wave
$\chi_b^{(-)}$, integrating over $b$'s spatial coordinate $\mathbf{r}_{bB}$
and internal coordinate $\zeta$.
Here and below, a bra of the form
$(\phi_0\,\chi_b^{(-)}|$ denotes this partial projection over
$b$'s center-of-mass and internal coordinates, leaving a ket in the
$x + A$ Hilbert space.
The result $|\rho_0\rangle$ is a state in the $x + A$ Hilbert space
depending on $\mathbf{r}_x$ and $\xi$.
In terms of $\rho_0$, the T-matrix becomes
$T_{0,c} = \langle \Psi_{xA}^c | \rho_0 \rangle$,
and the spectral identity
$\sum_c |\Psi_{xA}^c\rangle\langle\Psi_{xA}^c|\,
\delta(E_{x,0} - E^c)
= -\pi^{-1}\,\mathrm{Im}\,(E_{x,0}^+ - H_{xA})^{-1}$
yields
\begin{equation}
\frac{d^2\sigma}{dE_b\, d\Omega_b}\bigg|_{\mathrm{post}}
= -\frac{(2\pi)^4}{\pi v_a}\,
\mathrm{Im}\,
\langle \rho_0 |\, (E_{x,0}^+ - H_{xA})^{-1} \,| \rho_0 \rangle \,.
\label{eq:master_post}
\end{equation}
This is exact within the DWBA.
The propagator $(E_{x,0}^+ - H_{xA})^{-1}$ is the standard $x + A$
resolvent, identical to the one in the original IAV.
All nonspectator physics enters through the source $\rho_0$.

Since $V_{\mathrm{post}} = (V_{bx} + U_{bA} - U_{bB}) + (V_{bA} - U_{bA})$,
the source decomposes as
$|\rho_0\rangle = |\rho_0^{(\mathrm{sp})}\rangle
+ |\rho_0^{(\mathrm{nsp})}\rangle$, where the spectator part
\begin{equation}
|\rho_0^{(\mathrm{sp})}\rangle = ( \phi_0\, \chi_b^{(-)} |\,
(V_{bx} + U_{bA} - U_{bB}) \,| \chi_a^{(+)} \varphi_a \Phi_A \rangle
\label{eq:rho_sp}
\end{equation}
is the spectator source projected onto $b$'s ground state
(the $\mathcal{P}$-channel component of the IAV-type source),
and the nonspectator part
\begin{equation}
|\rho_0^{(\mathrm{nsp})}\rangle = ( \phi_0\, \chi_b^{(-)} |\,
(V_{bA} - U_{bA}) \,| \chi_a^{(+)} \varphi_a \Phi_A \rangle
\label{eq:rho_nsp}
\end{equation}
involves the correction $V_{bA} - U_{bA}$, the difference between
the true $b$-$A$ interaction and the optical model used in the IAV
framework.
Physically, this source receives contributions because the
projectile wave function $\varphi_a(\mathbf{r}_{bx}, \zeta)$ is a
superposition of $b$'s internal eigenstates,
$\varphi_a = \sum_n a_n(\mathbf{r}_{bx})\,\phi_n(\zeta)$.
The operator $V_{bA} - U_{bA}$ connects these components to the
detected ground state $\phi_0$ through its matrix elements
$\langle\phi_0|V_{bA}|\phi_n\rangle$: the diagonal element
($n = 0$) corrects for the difference between the true
ground-state interaction and the optical potential, while the
off-diagonal elements ($n \geq 1$) transfer amplitude from excited
components of $b$ within the projectile back to the ground state
during the interaction with the target.
A detailed decomposition is given in Sec.~\ref{sec:relation}.

The spectator source has a property essential for the optical reduction.
Because neither $V_{bx}$, $U_{bA}$, nor $U_{bB}$ depends on $\xi$, and
the only $\xi$-dependence in the integrand comes from $\Phi_A(\xi)$,
the result factorizes as
$\rho_0^{(\mathrm{sp})}(\mathbf{r}_x, \xi) = \tilde\rho_0(\mathbf{r}_x)
\,\Phi_A(\xi)$, where the reduced source
\begin{equation}
\tilde\rho_0(\mathbf{r}_x) = \int d\mathbf{r}_{bB}\, d\zeta\;
\chi_b^{(-)*}\,\phi_0^*\,
(V_{bx} + U_{bA} - U_{bB})\, \chi_a^{(+)}\, \varphi_a
\label{eq:rho_tilde}
\end{equation}
depends on $\mathbf{r}_x$ alone; here $U_{bA}$ is evaluated at
$\mathbf{r}_{bA} = \mathbf{r}_{bB}
+ \tfrac{m_x}{m_x + M_A}\,\mathbf{r}_x$ via the recoil relation while
the integration runs over $\mathbf{r}_{bB}$.
This factorization enables the Feshbach optical reduction
$\langle\Phi_A|(E_{x,0}^+ - H_{xA})^{-1}|\Phi_A\rangle
= G_x = (E_{x,0}^+ - E_A - K_x - U_x)^{-1}$,
where $U_x$ is the Feshbach optical potential for $x$ in the target
field.

The nonspectator source does not factorize in this way.
In coordinate representation,
\begin{align}
&\rho_0^{(\mathrm{nsp})}(\mathbf{r}_x, \xi)
= \int d\mathbf{r}_{bB}\, d\zeta\;
\chi_b^{(-)*}\,\phi_0^*
\notag \\
&\quad \times
[V_{bA}(\mathbf{r}_{bA}, \zeta, \xi) - U_{bA}(\mathbf{r}_{bA})]\,
\chi_a^{(+)}\,\varphi_a\,\Phi_A(\xi) \,.
\label{eq:rho_nsp_coord}
\end{align}
The product $V_{bA}\,\Phi_A(\xi)$ generates components on excited target
states, so $\rho_0^{(\mathrm{nsp})}$ is entangled in $\mathbf{r}_x$
and $\xi$ and the optical reduction does not apply.

Substituting into Eq.~(\ref{eq:master_post}), the cross section
separates into three terms,
\begin{equation}
\frac{d^2\sigma}{dE_b\, d\Omega_b}\bigg|_{\mathrm{post}}
= \sigma_{\mathrm{sp}} + \sigma_{\mathrm{cross}}
+ \sigma_{\mathrm{nsp}} \,.
\label{eq:three_terms}
\end{equation}
The spectator term is
\begin{equation}
\sigma_{\mathrm{sp}}
= -\frac{(2\pi)^4}{\pi v_a}\,
\mathrm{Im}\,\langle \tilde\rho_0 | G_x | \tilde\rho_0 \rangle \,,
\label{eq:sigma_sp}
\end{equation}
which has the same structure as Eq.~(2.11) of Ref.~\cite{Ichimura1985}.
The cross and nonspectator terms involve the full $x + A$ resolvent
$G_{\mathrm{full}} = (E_{x,0}^+ - H_{xA})^{-1}$ rather than the
optical Green's function $G_x$, because $\rho_0^{(\mathrm{nsp})}$
has non-trivial $\xi$-dependence:
\begin{align}
\sigma_{\mathrm{cross}}
&= -\frac{(2\pi)^4}{\pi v_a}\,
\mathrm{Im}\bigl[
\langle \rho_0^{(\mathrm{sp})} | G_{\mathrm{full}}
| \rho_0^{(\mathrm{nsp})} \rangle
+ (\mathrm{sp} \leftrightarrow \mathrm{nsp}) \bigr] \,,
\label{eq:sigma_cross} \\
\sigma_{\mathrm{nsp}}
&= -\frac{(2\pi)^4}{\pi v_a}\,
\mathrm{Im}\,\langle \rho_0^{(\mathrm{nsp})} | G_{\mathrm{full}}
| \rho_0^{(\mathrm{nsp})} \rangle \,.
\label{eq:sigma_nsp}
\end{align}
To evaluate these, one may expand $\rho_0^{(\mathrm{nsp})}$ in target
eigenstates as
$|\rho_0^{(\mathrm{nsp})}\rangle = \sum_\alpha |\Phi_\alpha\rangle
|\sigma_\alpha(\mathbf{r}_x)\rangle$,
leading to
$\mathrm{Im}\sum_{\alpha,\beta}
\langle\sigma_\alpha|[\mathbf{G}_x]_{\alpha\beta}|\sigma_\beta\rangle$,
where $[\mathbf{G}_x]_{\alpha\beta} = \langle\Phi_\alpha|
G_{\mathrm{full}}|\Phi_\beta\rangle$ is the coupled-channel Green's
function.
Equations~(\ref{eq:sigma_sp})--(\ref{eq:sigma_nsp}) constitute the
DWBA-exact result derived from the full Hamiltonian;
the cross and nonspectator terms require $G_{\mathrm{full}}$ and,
in general, coupled target channels, whereas an IAV-like
single-channel expression reappears only after an additional
simplifying assumption is made (see below).

The spectator term deserves a further remark regarding its
relationship to the standard IAV formula.
The present source $\tilde\rho_0$ [Eq.~(\ref{eq:rho_tilde})] involves
the projection $\langle\phi_0(\zeta)|$ and the composite projectile
wave function $\varphi_a(\mathbf{r}_{bx}, \zeta)$, whereas the
standard IAV source uses a structureless wave function
$\varphi_a^{\mathrm{sl}}(\mathbf{r}_{bx})$ with no $\zeta$ variable,
obtained by collapsing $b$'s internal Hilbert space to a single
degree of freedom.
The passage from $\tilde\rho_0$ to the corresponding structureless
source
$\tilde\rho_0^{\mathrm{sl}}(\mathbf{r}_x) = \int d\mathbf{r}_{bB}\,
\chi_b^{(-)*}(V_{bx}^{\mathrm{eff}} + U_{bA} - U_{bB})\chi_a^{(+)}
\varphi_a^{\mathrm{sl}}$
constitutes an additional \emph{structureless reduction} approximation,
distinct from the spectator replacement $V_{bA} \to U_{bA}$.
The full approximation chain from the exact Hamiltonian to the
standard IAV thus involves two steps:
(i) replacing $V_{bA}$ by $U_{bA}$ (spectator approximation), and
(ii) collapsing $b$'s internal space by identifying
$\langle\phi_0|\varphi_a\rangle_\zeta \to
\varphi_a^{\mathrm{sl}}$ and
$\langle\phi_0|V_{bx}|\varphi_a\rangle_\zeta \to
V_{bx}^{\mathrm{eff}}\,\varphi_a^{\mathrm{sl}}$
(structureless reduction).
The present formalism retains both $h_b$ and the full $V_{bA}$;
Eq.~(\ref{eq:sigma_sp}) matches the standard IAV form only after
the structureless reduction is also applied.

Following IAV, the spectator term is further decomposed into elastic
breakup (EBU) and nonelastic breakup (NEB) contributions using the
operator identity~\cite{Ichimura1985}
\begin{equation}
\mathrm{Im}\,G_x
= (1 + G_x^\dagger U_x^\dagger)\,\mathrm{Im}\,G_0\,(1 + U_x G_x)
+ G_x^\dagger\, W_x\, G_x \,,
\label{eq:ImGx}
\end{equation}
where $G_0 = (E_{x,0}^+ - E_A - K_x)^{-1}$ is the free Green's
function and $W_x = \mathrm{Im}\,U_x$ is the absorptive part of
the optical potential.
When inserted into Eq.~(\ref{eq:sigma_sp}), the first term on the
right gives the spectator elastic-breakup component, in which $x$
scatters elastically off $A$ while remaining in the continuum.
The second term gives the spectator nonelastic-breakup (breakup
fusion) component, in which $x$ is absorbed by the target.
Inserting Eq.~(\ref{eq:ImGx}) into Eq.~(\ref{eq:sigma_sp}) yields
the EBU/NEB decomposition of the spectator term,
\begin{equation}
\sigma_{\mathrm{sp}} = \sigma_{\mathrm{sp}}^{\mathrm{EBU}}
+ \sigma_{\mathrm{sp}}^{\mathrm{NEB}} \,.
\label{eq:sp_EBUNEB}
\end{equation}
The NEB part is
\begin{equation}
\sigma_{\mathrm{sp}}^{\mathrm{NEB}}
= -\frac{(2\pi)^4}{\pi v_a}\,
\langle \tilde\rho_0 | G_x^\dagger\, W_x\, G_x | \tilde\rho_0 \rangle \,.
\label{eq:sigma_sp_NEB}
\end{equation}
This has the same operator structure as the nonelastic breakup
formula computed by existing inclusive breakup
codes~\cite{Lei2015,Potel2015}
[Eq.~(2.21) of Ref.~\cite{Ichimura1985}]; the two coincide after
the structureless reduction discussed above.
The EBU
part $\sigma_{\mathrm{sp}}^{\mathrm{EBU}}$ is the spectator EBU
component generated by the first term of Eq.~(\ref{eq:ImGx}).

The spectator/nonspectator classification and the EBU/NEB
classification are logically distinct: the former concerns how the
source is generated, the latter concerns which $x + A$ final-state
channel that source feeds.
Once the source is modified by $V_{bA} - U_{bA}$, both the EBU and NEB
projections of the observed intact-$b$ cross section can change.
At the level of the full DWBA result, both elastic and nonelastic
breakup receive spectator, cross, and nonspectator contributions:
\begin{equation}
\sigma^{\mathrm{EBU}}
= \sigma_{\mathrm{sp}}^{\mathrm{EBU}}
+ \sigma_{\mathrm{cross}}^{\mathrm{EBU}}
+ \sigma_{\mathrm{nsp}}^{\mathrm{EBU}}
\label{eq:EBU_full_split}
\end{equation}
and
\begin{equation}
\sigma^{\mathrm{NEB}}
= \sigma_{\mathrm{sp}}^{\mathrm{NEB}}
+ \sigma_{\mathrm{cross}}^{\mathrm{NEB}}
+ \sigma_{\mathrm{nsp}}^{\mathrm{NEB}} \,.
\label{eq:NEB_full_split}
\end{equation}
For elastic breakup, $\sigma^{\mathrm{EBU}}$ corresponds to the
exclusive final channel $b(\phi_0)+x+A(\Phi_A)$, which can be fed
by both the spectator and nonspectator sources together with their
interference.
For nonelastic breakup, $\sigma^{\mathrm{NEB}}$ likewise receives all
three contributions; because NEB is the quantity that the IAV
formalism is designed to provide, the nonspectator correction is
most consequential here.
An EBU/NEB decomposition analogous to Eq.~(\ref{eq:ImGx}) can be
performed for each matrix element of the coupled-channel Green's
function $[\mathbf{G}_x]_{\alpha\beta}$, but it involves inelastic
target channels and does not reduce to the single-channel form
of Eq.~(\ref{eq:sigma_sp_NEB}).
In practice, the spectator elastic-breakup contribution
$\sigma_{\mathrm{sp}}^{\mathrm{EBU}}$ is typically computed
separately via the continuum-discretized coupled-channels (CDCC)
method~\cite{Kamimura1986,Austern1987,Yahiro2012,Thompson1988} or
direct DWBA, independent of the IAV sum rule.

The additional simplification is to construct $V_{bA}$ from nucleon
optical potentials~\cite{Koning2003} and neglect any explicit
$\xi$-dependence.
Then the nonspectator source is again proportional to $\Phi_A(\xi)$,
just like the spectator source, so the same Feshbach optical reduction
applies to all three terms.
Defining the nonspectator reduced source as
\begin{equation}
\tilde\rho_0^{(\mathrm{nsp})}(\mathbf{r}_x) = \int d\mathbf{r}_{bB}\,
d\zeta\;\chi_b^{(-)*}\,\phi_0^*\,
(V_{bA} - U_{bA})\,\chi_a^{(+)}\,\varphi_a \,,
\label{eq:rho_nsp_reduced}
\end{equation}
the total inclusive cross section in this restricted limit takes the form
\begin{equation}
\frac{d^2\sigma}{dE_b\, d\Omega_b}\bigg|_{\mathrm{post}}
= -\frac{(2\pi)^4}{\pi v_a}\,
\mathrm{Im}\,\langle \tilde\rho_0^{\mathrm{tot}} | G_x
| \tilde\rho_0^{\mathrm{tot}} \rangle \,,
\label{eq:master_level1}
\end{equation}
where $\tilde\rho_0^{\mathrm{tot}} = \tilde\rho_0 +
\tilde\rho_0^{(\mathrm{nsp})}$.
In this restricted single-channel limit, the EBU/NEB decomposition of
Eq.~(\ref{eq:ImGx}) applies to the full cross section, and the NEB part
becomes
\begin{equation}
\sigma^{\mathrm{NEB}}
= -\frac{(2\pi)^4}{\pi v_a}\,
\langle \tilde\rho_0^{\mathrm{tot}} | G_x^\dagger\, W_x\, G_x
| \tilde\rho_0^{\mathrm{tot}} \rangle \,.
\label{eq:sigma_NEB_total}
\end{equation}
This is the central \emph{practical} result, not the general DWBA result.
It states that, once the target dependence of $V_{bA}$ has been
suppressed, the nonelastic breakup cross section has the same operator
structure as Eq.~(\ref{eq:sigma_sp_NEB}), with the source
$\tilde\rho_0$ replaced by the total source
$\tilde\rho_0^{\mathrm{tot}} = \tilde\rho_0 +
\tilde\rho_0^{(\mathrm{nsp})}$.
After the additional structureless reduction discussed above, this
becomes the standard IAV expression with a modified source.
In existing IAV codes that compute
$\langle\tilde\rho_0|G_x^\dagger W_x G_x|\tilde\rho_0\rangle$,
the propagator and NEB projection remain unchanged; the extension
consists of replacing $\tilde\rho_0$ by $\tilde\rho_0^{\mathrm{tot}}$.
Constructing the new source $\tilde\rho_0^{(\mathrm{nsp})}$ is itself
a nontrivial step, discussed in Sec.~\ref{sec:discussion}.

\section{Prior-form sum rule}
\label{sec:prior}

While the post-form derivation provides the most transparent view of
how the spectator approximation enters and how it is removed, practical
implementations of the IAV formalism use the prior
form~\cite{Potel2015,Lei2015} because the post-form matrix element
suffers from a convergence problem~\cite{Ichimura1985,Lei2025post}.
The entrance-channel Hamiltonian is
$H_{\mathrm{ent}} = H_A + K_a + U_a + h_b + K_{bx} + V_{bx}$,
where $K_b + K_x = K_a + K_{bx}$ by Jacobi transformation and $U_a$
is the entrance-channel optical potential.
The prior-form residual interaction is
\begin{equation}
V_{\mathrm{prior}} = H - H_{\mathrm{ent}} = V_{xA} + V_{bA} - U_a \,.
\label{eq:Vprior}
\end{equation}
In the standard IAV with $V_{bA} \to U_{bA}$, this reduces to
$V_{\mathrm{prior}}^{\mathrm{IAV}} = V_{xA} + U_{bA} - U_a$.

The prior-form DWBA T-matrix
$T_{0,c}' = \langle \chi_b^{(-)} \phi_0\, \Psi_{xA}^c |\,
V_{\mathrm{prior}} \,| \chi_a^{(+)} \varphi_a \Phi_A \rangle$
defines a prior source
$|\rho_0'\rangle = ( \phi_0\, \chi_b^{(-)}|\,
V_{\mathrm{prior}} \,| \chi_a^{(+)} \varphi_a \Phi_A \rangle$
whose contribution to the NEB cross section is the Udagawa-Tamura
(UT) term~\cite{Udagawa1981,Lei2018prior}.
Because the exit-channel optical potential $U_{bB}$ is complex,
the prior-form DWBA does not coincide with the post-form result;
the difference is made up by the nonorthogonality (NO) and
interference (IN) corrections identified in
Ref.~\cite{Lei2018prior}.
The full IAV-equivalent prior-form NEB is
$\sigma^{\mathrm{NEB}} = \sigma_{\mathrm{UT}}^{\mathrm{NEB}}
+ \sigma_{\mathrm{NO}}^{\mathrm{NEB}}
+ \sigma_{\mathrm{IN}}^{\mathrm{NEB}}$,
and the nonspectator correction modifies both the UT and IN terms.
The NO term depends only on the channel overlap
$\psi_0^{\mathrm{NO}} = (\phi_0\,\chi_b^{(-)}|
\chi_a^{(+)}\varphi_a)$, which involves no interaction and is
therefore independent of the spectator approximation.
The IN term couples $\psi_0^{\mathrm{NO}}$ to the UT source and
is therefore also affected by the nonspectator correction.

I now restrict to the same reduced single-channel limit used in
Sec.~\ref{sec:post}, in which the target dependence of $V_{bA}$ is
suppressed and the projection onto the target ground state
introduces the optical propagator
$G_x = (E_{x,0}^+ - E_A - K_x - U_x)^{-1}$.
In this limit, the reduced spectator UT source is built from the
operator $U_x + U_{bA} - U_a$.
Since none of these potentials depends on the target coordinates
$\xi$, the spectator UT source factorizes in target space, just as
in the post form.
The reduced UT source is
\begin{equation}
\tilde\rho_0'^{\,\mathrm{UT}}(\mathbf{r}_x) = \int d\mathbf{r}_{bB}\,
d\zeta\;\chi_b^{(-)*}\,\phi_0^*\,
(U_x + U_{bA} - U_a)\,\chi_a^{(+)}\,\varphi_a \,,
\label{eq:rho_UT}
\end{equation}
and the spectator UT NEB cross section
is~\cite{Udagawa1981,Li1984,Lei2018prior}
\begin{equation}
\sigma_{\mathrm{UT}}^{\mathrm{NEB}}
= -\frac{(2\pi)^4}{\pi v_a}\,
\langle \tilde\rho_0'^{\,\mathrm{UT}} | G_x^\dagger\, W_x\, G_x
| \tilde\rho_0'^{\,\mathrm{UT}} \rangle \,,
\label{eq:sigma_UT}
\end{equation}
which has the same single-channel operator structure as the
post-form NEB [Eq.~(\ref{eq:sigma_sp_NEB})].
The standard IAV prior-form result is recovered after the
structureless reduction
$\langle\phi_0|\varphi_a\rangle_\zeta \to
\varphi_a^{\mathrm{sl}}$.

The nonspectator correction enters by replacing $U_{bA}$ with
$V_{bA}$ in the UT source, giving the total reduced prior source
\begin{equation}
\tilde\rho_0'^{\,\mathrm{tot}}(\mathbf{r}_x)
= \int d\mathbf{r}_{bB}\,d\zeta\;\chi_b^{(-)*}\,\phi_0^*\,
(U_x + V_{bA} - U_a)\,\chi_a^{(+)}\,\varphi_a \,,
\label{eq:rho_prior_tot}
\end{equation}
which decomposes as
$\tilde\rho_0'^{\,\mathrm{tot}} = \tilde\rho_0'^{\,\mathrm{UT}}
+ \tilde\rho_0'^{\,(\mathrm{nsp})}$, where the nonspectator
prior source
\begin{equation}
\tilde\rho_0'^{\,(\mathrm{nsp})}(\mathbf{r}_x)
= \int d\mathbf{r}_{bB}\,d\zeta\;\chi_b^{(-)*}\,\phi_0^*\,
(V_{bA} - U_{bA})\,\chi_a^{(+)}\,\varphi_a
\label{eq:rho_prior_nsp}
\end{equation}
involves the same operator $V_{bA} - U_{bA}$ as in the post form
[Eq.~(\ref{eq:rho_nsp})].
In both forms, the nonspectator correction is controlled by the
single operator $V_{bA} - U_{bA}$.
The total prior-form UT NEB cross section, including the
nonspectator correction, is therefore
\begin{equation}
\sigma_{\mathrm{UT}}^{\mathrm{NEB,tot}}
= -\frac{(2\pi)^4}{\pi v_a}\,
\langle \tilde\rho_0'^{\,\mathrm{tot}} | G_x^\dagger\, W_x\, G_x
| \tilde\rho_0'^{\,\mathrm{tot}} \rangle \,,
\label{eq:sigma_UT_tot}
\end{equation}
which has the same structure as the post-form result
[Eq.~(\ref{eq:sigma_NEB_total})] with the source
$\tilde\rho_0^{\mathrm{tot}}$ replaced by
$\tilde\rho_0'^{\,\mathrm{tot}}$.

The NO and IN corrections that complete the prior-form NEB involve
the nonorthogonality (NO) overlap between entrance and exit channels,
projected onto $b$'s ground state:
\begin{equation}
\psi_0^{\mathrm{NO}}(\mathbf{r}_x) = \int d\mathbf{r}_{bB}\, d\zeta\;
\chi_b^{(-)*}(\mathbf{r}_{bB})\,\phi_0^*(\zeta)\,
\chi_a^{(+)}(\mathbf{r}_a)\,\varphi_a(\mathbf{r}_{bx}, \zeta) \,.
\label{eq:psi_NO}
\end{equation}
This overlap does not involve any interaction and is therefore
independent of the spectator approximation.
The NO term is the Hussein-McVoy (HM)
contribution~\cite{Hussein1985}.
The NO and IN contributions to the NEB cross section
are~\cite{Hussein1985,Ichimura1985,Lei2018prior}
\begin{align}
\sigma_{\mathrm{NO}}^{\mathrm{NEB}}
&= -\frac{(2\pi)^4}{\pi v_a}\,
\langle \psi_0^{\mathrm{NO}} | W_x | \psi_0^{\mathrm{NO}} \rangle \,,
\label{eq:sigma_NO} \\
\sigma_{\mathrm{IN}}^{\mathrm{NEB}}
&= -\frac{(2\pi)^4}{\pi v_a}\,
2\,\mathrm{Re}\,
\langle \psi_0^{\mathrm{NO}} | W_x\, G_x
| \tilde\rho_0'^{\,\mathrm{tot}} \rangle \,.
\label{eq:sigma_IN}
\end{align}
The NO term is purely spectator: it depends only on
$\psi_0^{\mathrm{NO}}$ and carries no $V_{bA}$ dependence.
The IN term, by contrast, contains the total prior source
$\tilde\rho_0'^{\,\mathrm{tot}} = \tilde\rho_0'^{\,\mathrm{UT}}
+ \tilde\rho_0'^{\,(\mathrm{nsp})}$ and therefore decomposes into
spectator and nonspectator contributions.

\textit{Post-prior equivalence.} %
The difference of the post and prior residual interactions is
$V_{\mathrm{post}} - V_{\mathrm{prior}}
= V_{bx} - V_{xA} + U_a - U_{bB}$,
where $V_{bA}$ cancels exactly: whatever nonspectator effects
$V_{bA}$ introduces, it does so equally in both forms.
The post-prior identity that connects the two representations is
[cf.\ Eq.~(4.18) of Ref.~\cite{Ichimura1985}]
\begin{equation}
G_x\,|\tilde\rho_0'^{\,\mathrm{UT}}\rangle
= G_x\,|\tilde\rho_0\rangle - |\psi_0^{\mathrm{NO}}\rangle \,,
\label{eq:post_prior_identity}
\end{equation}
where $\tilde\rho_0$ is the post-form source
[Eq.~(\ref{eq:rho_tilde})].
Substituting this identity into the UT + NO + IN decomposition
and collecting terms recovers the post-form NEB result
[Eq.~(\ref{eq:sigma_sp_NEB})], establishing that
$\sigma_{\mathrm{UT}}^{\mathrm{NEB}}
+ \sigma_{\mathrm{NO}}^{\mathrm{NEB}}
+ \sigma_{\mathrm{IN}}^{\mathrm{NEB}}
= \sigma_{\mathrm{sp}}^{\mathrm{NEB}}$
for the spectator sector.
The identity~(\ref{eq:post_prior_identity}) connects only the
spectator UT and post sources.
The nonspectator reduced source
$\tilde\rho_0'^{\,(\mathrm{nsp})}$ [Eq.~(\ref{eq:rho_prior_nsp})]
is built from the same operator $V_{bA} - U_{bA}$ and the same
projection $\langle\phi_0\,\chi_b^{(-)}|\cdots|\chi_a^{(+)}
\varphi_a\rangle$ as the post-form nonspectator source
$\tilde\rho_0^{(\mathrm{nsp})}$
[Eq.~(\ref{eq:rho_nsp_reduced})], so
$\tilde\rho_0'^{\,(\mathrm{nsp})} = \tilde\rho_0^{(\mathrm{nsp})}$.
The identity therefore extends term by term:
$G_x|\tilde\rho_0'^{\,\mathrm{tot}}\rangle
= G_x|\tilde\rho_0^{\mathrm{tot}}\rangle
- |\psi_0^{\mathrm{NO}}\rangle$,
and the full (spectator + nonspectator) NEB cross section is the
same in both forms.
This equivalence is established here in the reduced single-channel
limit, where the optical propagator $G_x$ is common to both forms.
Its scope should not be confused with the classic post-prior
asymmetry of the standard sum rule: the prior form differs from the
post form by the nonorthogonality and interference terms
[Eqs.~(\ref{eq:sigma_NO}) and~(\ref{eq:sigma_IN})] regardless of the
spectator approximation, and the nonspectator correction
$V_{bA} - U_{bA}$, which cancels in
$V_{\mathrm{post}} - V_{\mathrm{prior}}$, enters both forms identically
and therefore introduces no additional post-prior terms of its own.

\section{Relation to the IAV formalism}
\label{sec:relation}

The spectator approximation can be understood through the Feshbach
projection formalism~\cite{Feshbach1958,Feshbach1962} in the
internal space of $b$.
There are two distinct questions here, and it is useful to keep them
separate from the outset.
The first is what the spectator approximation removes from the exact
composite-fragment problem.
The second is how the standard \emph{structureless} IAV observable
should be interpreted when it is embedded back into that composite
Hilbert space.
Defining the projectors $\mathcal{P} = |\phi_0\rangle\langle\phi_0|$
and $\mathcal{Q} = 1 - \mathcal{P}$ onto $b$'s ground state and
excited states, the full scattering state $|\Psi\rangle$ has both
$\mathcal{P}$ and $\mathcal{Q}$ components.
At the level of the underlying composite-fragment problem, the
experiment can in principle distinguish two classes of observables.
Detecting $b$ in its ground state probes the $\mathcal{P}$ channel.
Detecting excited states of $b$, or the breakup products of $b$,
probes the $\mathcal{Q}$ sector.
The exact formalism developed here computes the $\mathcal{P}$-channel
cross section, namely the observable in which the detected fragment is
projected onto $\phi_0$ as in Eq.~(\ref{eq:master_post}).
How the standard structureless IAV should be interpreted relative to
that distinction is a separate question, addressed below.

The spectator approximation replaces $V_{bA}(\mathbf{r}_{bA}, \zeta, \xi)$
by an optical potential $U_{bA}(\mathbf{r}_{bA})$ that is independent of
$b$'s internal coordinates $\zeta$ and the target coordinates $\xi$.
Because $U_{bA}$ does not depend on $\zeta$, it is diagonal in
$b$'s eigenstates:
$\langle\phi_0|U_{bA}|\phi_n\rangle = U_{bA}\,\delta_{n0}$.
This means that the off-diagonal matrix elements
$V_{0n}^{(bA)} = \langle\phi_0|V_{bA}|\phi_n\rangle$ for $n \geq 1$,
which connect the $\mathcal{Q}$-components of the projectile wave
function to the detected $\mathcal{P}$-state through the $b$-$A$
interaction, are discarded.
Note that the spectator approximation does not eliminate all
$\mathcal{Q} \to \mathcal{P}$ transitions: the binding interaction
$V_{bx}$ can still change $b$'s internal state (since it depends on
$\zeta$), and the corresponding matrix elements
$\langle\phi_0|V_{bx}|\phi_n\rangle$ are retained.
What is lost specifically is the role of $V_{bA}$ in mediating
transitions between $b$'s internal states during the interaction
with the target, and the $\xi$-dependence of the diagonal matrix
element $V_{00}^{(bA)}$ that couples $b$'s finite size to target
excitations.

The nonspectator generalization developed in this paper retains the
full $V_{bA}$ with all its $\zeta$- and $\xi$-dependence, including
both the diagonal and off-diagonal matrix elements in $b$'s internal
space.
The nonspectator source $\rho_0^{(\mathrm{nsp})}$
[Eq.~(\ref{eq:rho_nsp})] captures precisely the contributions that
the spectator approximation misses.
Expanding the projectile wave function in $b$'s eigenstates,
$\varphi_a(\mathbf{r}_{bx}, \zeta) = \sum_n a_n(\mathbf{r}_{bx})
\,\phi_n(\zeta)$, the nonspectator source receives contributions
from two distinct mechanisms:
\begin{align}
\langle\phi_0|(V_{bA} - U_{bA})|\varphi_a\rangle_\zeta
&= \bigl(V_{00}^{(bA)} - U_{bA}\bigr)\, a_0(\mathbf{r}_{bx})
\notag \\
&+ \sum_{n \geq 1} V_{0n}^{(bA)}\, a_n(\mathbf{r}_{bx}) \,,
\label{eq:nsp_two_mechanisms}
\end{align}
where $V_{0n}^{(bA)} = \langle\phi_0|V_{bA}|\phi_n\rangle_\zeta$.
The first term is the diagonal correction: $b$ remains in its
ground state ($\mathcal{P}$-space), but the true ground-state
interaction $V_{00}^{(bA)}(\mathbf{r}_{bA}, \xi)$ differs from the
optical potential $U_{bA}(\mathbf{r}_{bA})$ because it retains the
$\xi$-dependence and finite-size effects averaged out in $U_{bA}$.
The second term is the $\mathcal{Q}$-space contribution: $b$ is found
in an excited component $\phi_n$ (with $n \geq 1$) within the
projectile $a$, and the off-diagonal matrix element $V_{0n}^{(bA)}$
mediates the transition $\mathcal{Q} \to \mathcal{P}$ during the
interaction with the target.
This term represents the coherent return from $\mathcal{Q}$-space
to $\mathcal{P}$-space that the experiment detects but the spectator
approximation discards.
In the spectator limit $V_{bA} \to U_{bA}$, both terms vanish:
$V_{00}^{(bA)} \to U_{bA}$ and $V_{0n}^{(bA)} \to U_{bA}\delta_{n0}$,
recovering $\rho_0^{(\mathrm{nsp})} = 0$ and the IAV sum rule.

\textit{What observable does the structureless IAV represent?} %
Any identification of the structureless IAV with a composite-fragment
observable is necessarily approximate, requiring at minimum a closure
approximation ($G_{x,n} \approx G_x$, with the channel-dependent
propagator $G_{x,n}$ defined in Eq.~(\ref{eq:sigma_n}) below),
identical exit-channel
distorted waves and detector acceptance for all internal states $n$,
and the structureless reduction
$(\varphi_a, V_{bx}) \to (\varphi_a^{\mathrm{sl}},
V_{bx}^{\mathrm{eff}})$.
With these caveats stated at the outset, I now show that the
structureless IAV most naturally corresponds to the \emph{total}
inclusive cross section summed over all of $b$'s internal states,
rather than the state-resolved cross section for any individual
$\phi_n$.

The post-form master formula [Eq.~(\ref{eq:master_post})] and its
prior-form counterpart (which in the reduced single-channel
formulation becomes Eq.~(\ref{eq:sigma_UT})) describe the
state-resolved observable in which the detected fragment is projected
onto $\phi_0$.
The standard IAV formalism, by contrast, does not carry an explicit
projection onto any internal state of $b$, because $b$ has been made
structureless from the start.

The distinction between the state-resolved and total observables is
most transparent in the language of projection operators in $b$'s
internal space.
Working within the reduced single-channel limit where the explicit
target dependence has been projected out, define the unprojected source
$\mathcal{S}(\mathbf{r}_x, \zeta) = \langle\chi_b^{(-)}|
V_{\mathrm{post}}\,\chi_a^{(+)}\,\varphi_a\rangle_{\mathbf{r}_{bB}}$,
which retains $b$'s internal coordinate $\zeta$ through
$V_{bx}(\mathbf{r}_{bx}, \zeta)$, $V_{bA}(\mathbf{r}_{bA}, \zeta)$,
and $\varphi_a(\mathbf{r}_{bx}, \zeta)$; the target coordinate $\xi$ no
longer appears explicitly in $V_{bA}$ here because the target
dependence has been projected out in this reduced limit.
The state-resolved NEB cross section for detecting $b$ in internal
state $\phi_n$ projects this source with
$\mathcal{P}_n = |\phi_n\rangle\langle\phi_n|$:
\begin{equation}
\sigma_n^{\mathrm{NEB}} = -\frac{(2\pi)^4}{\pi v_a}\,
\langle\mathcal{S}|\,
\mathcal{P}_n \otimes G_{x,n}^\dagger W_{x,n} G_{x,n}\,
|\mathcal{S}\rangle \,,
\label{eq:sigma_n}
\end{equation}
where
$G_{x,n} = (E_{x,n}^+ - E_A - K_x - U_{x,n})^{-1}$
is the $x + A$ optical propagator and
$W_{x,n} = \mathrm{Im}\,U_{x,n}$ the corresponding absorptive potential,
both evaluated at the channel energy
$E_{x,n} = E - E_b - \varepsilon_n$ appropriate to detecting $b$ in its
internal state $\phi_n$.
For $n = 0$ they reduce to the ground-state quantities $G_x$
(defined following Eq.~(\ref{eq:rho_tilde})) and
$W_x = \mathrm{Im}\,U_x$ [Eq.~(\ref{eq:ImGx})] introduced above,
with $U_{x,0} \equiv U_x$ the optical potential evaluated at the
ground-state channel energy $E_{x,0}$.
If the internal excitation energies $\varepsilon_n - \varepsilon_0$ are
small compared to $E_{x,0}$, one may approximate
$G_{x,n} \approx G_x$ for all relevant $n$ (closure approximation).
The total NEB cross section then becomes, by completeness
$\sum_n \mathcal{P}_n = \mathbf{1}_\zeta$,
\begin{equation}
\sigma_{\mathrm{total}}^{\mathrm{NEB}}
= \sum_n \sigma_n^{\mathrm{NEB}}
\approx -\frac{(2\pi)^4}{\pi v_a}\,
\langle\mathcal{S}|\,
\mathbf{1}_\zeta \otimes G_x^\dagger W_x G_x\,
|\mathcal{S}\rangle \,.
\label{eq:sigma_total_trace}
\end{equation}
The projection $\mathcal{P}_0 = |\phi_0\rangle\langle\phi_0|$ selects
the ground-state channel $\sigma_0^{\mathrm{NEB}}$;
the identity $\mathbf{1}_\zeta$ sums over all channels, giving
$\sigma_{\mathrm{total}}^{\mathrm{NEB}}$.
These are the two limiting choices relevant to the present discussion.

The source-level projection $\mathcal{P}_0$ versus $\mathbf{1}_\zeta$
has a direct counterpart at the Hamiltonian level through the Feshbach
formalism.
The state-resolved cross section $\sigma_0^{\mathrm{NEB}}$ corresponds
to working in the $\mathcal{P}$-projected space.
In the Feshbach framework, the effective $\mathcal{P}$-space
Hamiltonian includes a polarization potential from virtual excitations
to $\mathcal{Q}$-space~\cite{Feshbach1958,Feshbach1962},
\begin{equation}
H_{\mathrm{eff}} = \mathcal{P}H\mathcal{P}
+ \mathcal{P}H\mathcal{Q}\,(E - \mathcal{Q}H\mathcal{Q})^{-1}\,
\mathcal{Q}H\mathcal{P} \,,
\label{eq:Heff_PQ}
\end{equation}
which accounts for the fact that, even when $b$ is detected in its
ground state, it may have been virtually excited to $\mathcal{Q}$-space
during the reaction.
The total cross section $\sigma_{\mathrm{total}}^{\mathrm{NEB}}$,
by contrast, sums over all $\mathcal{P}$ and $\mathcal{Q}$
final states and therefore uses the \emph{full} Hamiltonian $H$
without any Feshbach reduction.

The standard structureless IAV has no $\zeta$ degree of freedom
and no $\mathcal{P}/\mathcal{Q}$ decomposition.
In the composite language, it does not impose a projection onto any
particular internal state of $b$.
Within the composite framework, this is the analog of using
$\mathbf{1}_\zeta = \mathcal{P} + \mathcal{Q}$ (the trace over
all internal states) rather than the ground-state projection
$\mathcal{P}_0$.
The structureless IAV is therefore most naturally interpreted as an
approximation to $\sigma_{\mathrm{total}}^{\mathrm{NEB}}$, not to the
state-resolved $\sigma_0^{\mathrm{NEB}}$.
The difference between the two is the $\mathcal{Q}$-space
contribution: virtual excitations of $b$ that feed back into the
detected ground-state channel through $H_{\mathrm{eff}}$ but are
included automatically when no ground-state projection is performed.
Subject to the closure and structureless-reduction approximations
stated above, the structureless IAV is therefore associated with the
total cross section,
\begin{equation}
\sigma_{\mathrm{IAV}}^{\mathrm{NEB}}(\text{structureless } b)
\;\longleftrightarrow\;
\sigma_{\mathrm{total}}^{\mathrm{NEB}}
= \sum_n \sigma_n^{\mathrm{NEB}} \,,
\label{eq:IAV_is_total}
\end{equation}
rather than the state-resolved cross section $\sigma_0^{\mathrm{NEB}}$
for detecting $b$ in its ground state.
This correspondence is subject to several caveats beyond the closure
approximation itself.
First, the exit-channel distorted wave $\chi_b^{(-)}$ and the
detector acceptance are assumed identical for all internal states $n$;
in practice, the kinetic energy $E_b$ and angular distribution of $b$
differ between the ground state and excited/continuum states.
Second, for $n \geq 1$ (excited or broken-up $b$), the experimental
observable is the detection of $b$'s breakup products, which is a
different final-state selection from detecting an intact $b$.
Third, the quantitative correspondence depends on how accurately
the structureless wave functions and interactions approximate the
$\zeta$-traced composite quantities, which involves the additional
structureless reduction discussed in Sec.~\ref{sec:post}.
Equation~(\ref{eq:IAV_is_total}) should therefore be understood as a
conceptual correspondence valid under these approximations, not as an
exact identity.

For the deuteron, the closure approximation is qualitatively valid:
the only bound state is the ground state ($\varepsilon_0 = -2.224$~MeV),
and the low-lying continuum states that carry most of the spectral
weight of $\varphi_a$ have excitation energies of a few MeV, modest
compared to typical reaction energies $E_{x,0} \sim 10$--$30$~MeV.
The approximation is least controlled near the Coulomb barrier, where
$E_{x,0}$ is smallest and $W_x$ varies rapidly with energy; there the
channel-dependent propagator $G_{x,n}$ should be retained rather than
replaced by $G_x$.

This observation has two distinct implications.
At the \emph{formalism} level, the nonspectator generalization
provides corrections to the IAV source through $V_{bA} - U_{bA}$,
improving the dynamical content of the calculation.
At the \emph{observable} level, the present formalism computes
$\sigma_0^{\mathrm{NEB}}$ (detecting $b$ in its ground state),
whereas the structureless IAV approximates
$\sigma_{\mathrm{total}}^{\mathrm{NEB}}$ (summed over all internal
states of $b$).
These are in general different observables, and the nonspectator
formalism does not merely ``correct'' the IAV result but accesses
a state-resolved quantity that the structureless theory cannot
distinguish.
The $\mathcal{Q}$-channel cross sections
$\sigma_n^{\mathrm{NEB}}$ ($n \geq 1$), corresponding to detection
of $b$'s breakup products, can be obtained from the same formalism
by replacing $\langle\phi_0|$ with $\langle\phi_n|$ in the source
function, with the obvious extension to continuum-normalized states
or discretized bins for $n$ in the continuum.

The EBU/NEB decomposition of the original IAV applies to
$\sigma_{\mathrm{sp}}$ through
$\mathrm{Im}\,G_x = G_x^\dagger W_x G_x + (\mathrm{EB})$.
Accordingly, once the full source is considered, both EBU and NEB are
organized as in Eqs.~(\ref{eq:EBU_full_split}) and
(\ref{eq:NEB_full_split}).
The quantities $\sigma_{\mathrm{sp}}^{\mathrm{EBU}}$ and
$\sigma_{\mathrm{sp}}^{\mathrm{NEB}}$ are only the spectator parts of
those full observables.
For elastic breakup, the full quantity $\sigma^{\mathrm{EBU}}$
corresponds to the exclusive final channel $b(\phi_0)+x+A(\Phi_A)$.
In other words, nonspectator dynamics should not be identified with
NEB alone.
It modifies the source before the final-state projection, and the same
modified source can feed either the EBU or the NEB sector.
The essential point is that ``spectator versus nonspectator'' is a
source-level classification, whereas ``EBU versus NEB'' is a
final-state classification.
For the nonspectator terms, an analogous decomposition exists via the
spectral representation of $G_{\mathrm{full}}$ but involves inelastic
target channels.

Within the DWBA, all nonspectator effects enter through the source:
at the level of the exact master formula
[Eq.~(\ref{eq:master_post})], the propagator
$(E_{x,0}^+ - H_{xA})^{-1}$ is the same regardless of whether the
spectator approximation is made.
(In the reduced single-channel limit, this common propagator becomes
$G_x$ for both the spectator and nonspectator contributions.)
A beyond-DWBA treatment replaces the entrance-channel distorted wave
$\chi_a^{(+)}\varphi_a$ by a coupled-channel solution such as the
CDCC wave function~\cite{Lei2023cdcc}, modifying the source while
leaving the propagator structure unchanged.
The present nonspectator correction to the source applies equally
in that framework.

\section{Tidal estimate for the deuteron}
\label{sec:estimate}

\begin{figure*}[t]
\centering
\includegraphics[width=0.96\textwidth]{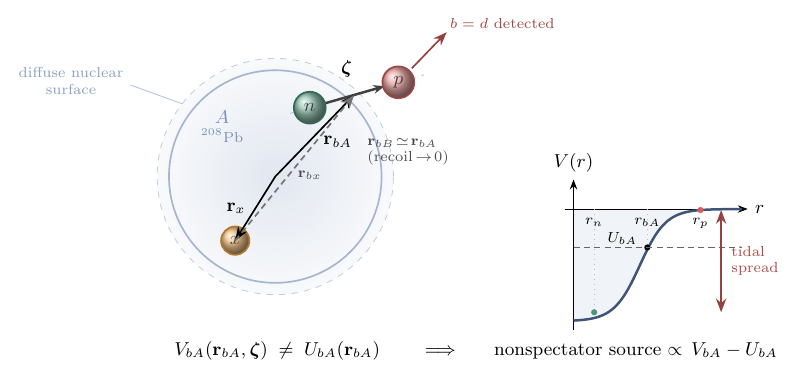}
\caption{\label{fig:tidal}
Schematic of the nonspectator (tidal) mechanism, illustrated for
$b = d$ on ${}^{208}\mathrm{Pb}$.
Left: in the exit channel the detected deuteron $b = p + n$ grazes
the diffuse nuclear surface, with center-of-mass position
$\mathbf{r}_{bA}$ and internal ($p$-$n$) coordinate
$\boldsymbol{\zeta}$ drawn as a vector from $n$ to $p$; the unobserved
fragment $x$ is absorbed by the target. The Jacobi coordinates used
in the text are indicated, with $\mathbf{r}_{bB} \simeq \mathbf{r}_{bA}$
up to the residual recoil.
Right: the $b$-$A$ potential $V(r)$ sampled along the line through the
constituents. The spectator approximation replaces the true interaction
by the optical potential $U_{bA}$, the potential a structureless (point)
$b$ would feel at its center of mass $r_{bA}$ (dashed line). Because the
proton (outer, $r_p$) and neutron (inner, $r_n$) sit at different radii,
they sample $V$ at different depths; this spread is the tidal variation
that $U_{bA}$ averages out, and it drives the nonspectator source through
the residual operator $V_{bA} - U_{bA}$.}
\end{figure*}

To assess the magnitude of the operator $V_{bA} - U_{bA}$
(illustrated in Fig.~\ref{fig:tidal}), I consider
$b = d$ (deuteron) on ${}^{208}\mathrm{Pb}$.
For the deuteron I write $\mathbf{r}_d \equiv \mathbf{r}_{bA}$ for the
$b$-$A$ separation introduced in Sec.~\ref{sec:post}.
The deuteron-target interaction for a two-body $p$-$n$ system with
internal coordinate $\boldsymbol\zeta$ is
\begin{equation}
V_{dA}(\mathbf{r}_d, \boldsymbol\zeta, \xi)
= V_{pA}\bigl(\mathbf{r}_d + \tfrac{1}{2}\boldsymbol\zeta,\, \xi\bigr)
+ V_{nA}\bigl(\mathbf{r}_d - \tfrac{1}{2}\boldsymbol\zeta,\, \xi\bigr) \,.
\label{eq:VdA}
\end{equation}
A natural reference potential is the adiabatic folding
potential~\cite{Johnson1970,Harvey1971},
defined by averaging over the deuteron ground state,
\begin{equation}
U_d^{\mathrm{fold}}(\mathbf{r}_d) = \int |\phi_d(\boldsymbol\zeta)|^2\,
[V_{pA} + V_{nA}]\, d\boldsymbol\zeta \,.
\label{eq:Ufold}
\end{equation}
This choice ensures that the diagonal nonspectator correction
$\langle\phi_d|(V_{dA} - U_d^{\mathrm{fold}})|\phi_d\rangle$
vanishes by construction, isolating the off-diagonal (tidal)
contributions.
The dependence on the choice of reference potential is discussed
in Sec.~\ref{sec:discussion}.

Expanding $V_{dA}$ in a Taylor series of $\boldsymbol\zeta$ about
$\mathbf{r}_d$ gives
\begin{align}
V_{dA} &= \Sigma V(\mathbf{r}_d, \xi)
+ \boldsymbol\zeta \cdot \nabla_{\mathbf{r}_d}\,\delta V
\notag \\
&+ \frac{1}{8}\,\zeta_i \zeta_j\,
\partial_i \partial_j\, \Sigma V
+ \cdots \,,
\label{eq:multipole}
\end{align}
where $\Sigma V = V_{pA} + V_{nA}$ is the isoscalar sum and
$\delta V = (V_{pA} - V_{nA})/2$ is the isovector difference,
both evaluated at $\mathbf{r}_d$.
The zeroth-order (monopole) term $\Sigma V(\mathbf{r}_d)$ is the
zero-range limit of the folding potential; the true folding
potential [Eq.~(\ref{eq:Ufold})] includes finite-size corrections:
for an $s$-wave deuteron,
\begin{equation}
U_d^{\mathrm{fold}} = \Sigma V
+ \frac{\langle\zeta^2\rangle}{24}\,\nabla^2 \Sigma V + \cdots \,,
\label{eq:Ufold_expand}
\end{equation}
where $\langle\zeta^2\rangle = \int |\phi_d|^2\,\zeta^2\,
d\boldsymbol\zeta$ and the dipole average
$\langle\boldsymbol\zeta\rangle = 0$ by parity.
The nonspectator operator is therefore
\begin{align}
V_{dA} - U_d^{\mathrm{fold}}
&= \boldsymbol\zeta \cdot \nabla\,\delta V
\notag \\
&+ \frac{1}{8}\biggl(\zeta_i\zeta_j
- \frac{\delta_{ij}}{3}\langle\zeta^2\rangle\biggr)
\partial_i\partial_j\,\Sigma V + \cdots \,.
\label{eq:VdA_minus_Ufold}
\end{align}
The first (dipole) term couples the deuteron's $E1$ response
to the gradient of the \emph{isovector} potential $\delta V$;
it enters the nonspectator source through off-diagonal matrix
elements $\langle\phi_0|\boldsymbol\zeta|\phi_n\rangle$.
The second (quadrupole) term couples the $E2$ response to the
curvature of the isoscalar potential $\Sigma V$.
Each multipole provides a systematic truncation for the
coupled-channel evaluation of the nonspectator terms.

Since the diagonal matrix element vanishes with this choice of
$U_{bA}$, a useful diagnostic is the rms
fluctuation
$\Delta V_{\mathrm{tidal}}(r_d)
= \sqrt{\langle\phi_d|(V_{dA} - U_d^{\mathrm{fold}})^2
|\phi_d\rangle}$.
Because the diagonal element vanishes by the folding choice,
$\Delta V_{\mathrm{tidal}}$ is, by closure, the root-sum-square of all
off-diagonal couplings
$\langle\phi_0|(V_{dA} - U_d^{\mathrm{fold}})|\phi_n\rangle$; it is thus
the aggregate strength that actually feeds the nonspectator source,
not merely a diagonal expectation value.
For a back-of-the-envelope estimate, the dominant contribution
at the nuclear surface comes from the dipole term.
The isovector nuclear potential for ${}^{208}\mathrm{Pb}$ has a
typical strength of $|V_p - V_n|/2 \approx 5$--$8$~MeV
(predominantly the symmetry/Lane potential), which changes over a
surface diffuseness of $\sim 0.7$~fm, giving an isovector gradient
$|\nabla\,\delta V| \sim 10$~MeV/fm.
For the Coulomb contribution, the gradient at the nuclear surface
($r \approx 7$~fm) is $|\nabla V_C|/(2) \approx
Z e^2/(2 r^2) \approx 1.5$~MeV/fm, which is subdominant.
With the deuteron's rms $p$-$n$ separation
$\sqrt{\langle\zeta^2\rangle} \approx 3.9$~fm, the dipole tidal
correction is of order
$|\nabla\,\delta V| \times \zeta_{\mathrm{rms}}/\sqrt{3}
\sim 10 \times 2.25 \approx 22$~MeV,
where the factor $1/\sqrt{3}$ arises from the angular average
of $\boldsymbol\zeta \cdot \hat{\nabla}$ over the $s$-wave
deuteron.
While this is smaller than the depth of $V_{dA}$ itself, it is
comparable to or larger than the deuteron binding energy
($\epsilon_d = 2.224$~MeV) and is not a small perturbation.
The quadrupole contribution involves the traceless combination
$(\zeta_i\zeta_j - \delta_{ij}\langle\zeta^2\rangle/3)$ coupled
to $\partial_i\partial_j\,\Sigma V$.
For a Woods-Saxon potential with depth $V_0 \sim 50$~MeV,
radius $R \approx 7$~fm, and diffuseness $a \approx 0.65$~fm,
the isoscalar Laplacian at the nuclear surface is dominated by
the $2V'/r$ contribution (the radial second derivative $V''$
vanishes at $r = R$ for a Woods-Saxon shape), giving
$|\nabla^2 \Sigma V| \sim V_0/(aR) \sim 11~\mathrm{MeV/fm}^2$.
With $\langle\zeta^2\rangle \approx 15~\mathrm{fm}^2$, the
quadrupole scale is
$\langle\zeta^2\rangle\,|\nabla^2\Sigma V|/8 \sim 21$~MeV,
comparable to the dipole estimate.
This uses the scalar Laplacian as a dimensional scale; the traceless
quadrupole operator in fact contracts with the anisotropy
$V'' - V'/r$ of the Hessian rather than its trace
$V'' + 2V'/r$, so the value is an order-of-magnitude bound rather than
the precise quadrupole matrix element.
The relevant quantity for the nonspectator source is the
off-diagonal transition matrix element
$\langle\phi_0|Q_{ij}|\phi_n\rangle$, with
$Q_{ij} = \zeta_i\zeta_j - \delta_{ij}\langle\zeta^2\rangle/3$,
which is reduced by the rapid falloff of the deuteron wave function
at large $\zeta$.
These estimates indicate that the nonspectator operator is not a
small perturbation at the nuclear surface.
Whether this operator-level scale translates into a significant
difference between $\sigma_0^{\mathrm{NEB}}$ and
$\sigma_{\mathrm{total}}^{\mathrm{NEB}}$ requires numerical
evaluation of the source integrals.

\section{Discussion}
\label{sec:discussion}

The main results of this work are summarized in Table~\ref{tab:summary},
which compares the standard IAV formalism with the nonspectator
generalization.
The central finding is that, within the DWBA, all nonspectator
effects are encoded in the operator $V_{bA} - U_{bA}$ acting in the source
function, while the propagator remains unchanged.
Because this operator modifies the source before the final-state
projection, it can in principle change both the elastic-breakup and
nonelastic-breakup components of the observed intact-$b$ cross section,
even though the practical IAV machinery is primarily aimed at NEB.
This operator has a physical interpretation as the tidal correction
across $b$'s internal extent (Fig.~\ref{fig:tidal}): the individual
constituents of $b$
experience different nuclear potentials at their respective positions,
and this spatial variation is averaged out in the optical potential $U_{bA}$.
For compact fragments such as the alpha particle
($\epsilon_\alpha = 28.3$~MeV, $r_{\mathrm{rms}} \approx 1.5$~fm),
the tidal correction is small and the spectator approximation is
accurate.
For the deuteron ($\epsilon_d = 2.224$~MeV, rms $p$-$n$
separation $\approx 3.9$~fm), the tidal estimate of
Sec.~\ref{sec:estimate} shows that the nonspectator operator is of
order several tens of MeV at the nuclear surface, far exceeding the
deuteron binding energy and not a small perturbation; whether this
operator-level scale produces a significant change in the cross section
awaits numerical evaluation of the source integrals.

\begin{table*}[t]
\caption{Comparison of the standard IAV formalism (spectator
approximation) with the nonspectator generalization derived in this
work. Post-prior equivalence is listed for the reduced single-channel
formulation, with the nonorthogonality and interference terms
completing the prior form.}
\label{tab:summary}
\begin{ruledtabular}
\begin{tabular}{lcc}
Quantity & IAV & Nonspectator \\
\hline
$b$ structure & Structureless & Composite \\
$V_{\mathrm{post}}$ & $V_{bx} + U_{bA} - U_{bB}$ & $V_{bx} + V_{bA} - U_{bB}$ \\
Internal-state treatment & No explicit internal-state projection &
  Explicit $\langle\phi_0|$ projection \\
Observable (conceptual) & $\sigma_{\mathrm{total}}^{\mathrm{NEB}}$ (under closure, structureless reduction) &
  $\sigma_0^{\mathrm{NEB}}$ \\
Post-prior & Yes & Yes \\
\end{tabular}
\end{ruledtabular}
\end{table*}

The regime where nonspectator effects are expected to be most
important includes low-energy reactions
($E_a \lesssim 30$~MeV/nucleon), where the projectile spends
significant time in the nuclear field;
heavy targets (large $A$), where the nuclear potential varies strongly
over the deuteron's spatial extent;
and reactions near the Coulomb barrier, where stripping and transfer
processes compete with direct breakup.
At high energies ($E_a > 100$~MeV/nucleon), the eikonal approximation
becomes valid, the deuteron traverses the nuclear field rapidly, and
nonspectator effects are expected to be
suppressed~\cite{Serber1947}.
Notably, a related effect has been identified at intermediate
energies in nucleon-knockout reactions~\cite{Lei2026}, where a
double Feshbach projection of the three-body Hamiltonian reveals
that the standard additive model for composite projectiles misses
induced interactions from excluded projectile configurations,
leading to systematic overestimates of the stripping cross section
and apparent spectroscopic quenching.
The underlying physics is the same: treating a composite fragment
as structureless neglects terms that can significantly modify the
cross section.

For practical implementation, it is again useful to distinguish the
simplified and fully microscopic levels.
In the simplified scheme where $V_{bA}$ is built from nucleon optical
potentials and does not depend explicitly on $\xi$, the calculation
remains close to existing IAV machinery, but one should distinguish the
post and prior representations.
In post form, one replaces the standard source $\tilde\rho_0$ by
$\tilde\rho_0^{\mathrm{tot}} = \tilde\rho_0 +
\tilde\rho_0^{(\mathrm{nsp})}$ in Eq.~(\ref{eq:sigma_NEB_total}).
In prior form, which is what practical codes usually implement, the
analogous extension is to add the reduced nonspectator source
$\tilde\rho_0'^{(\mathrm{nsp})}$ to the UT/HM building blocks while
retaining the same optical propagation.
The genuinely new step in either representation is the evaluation of
the reduced nonspectator source, which requires integration over $b$'s
internal coordinate $\zeta$ with the coordinate-dependent operator
$V_{bA}(\mathbf{r}_{bA}, \zeta) - U_{bA}(\mathbf{r}_{bA})$.
For the deuteron,
$V_{bA} = V_{pA}(\mathbf{r}_{bA} + \boldsymbol\zeta/2) +
V_{nA}(\mathbf{r}_{bA} - \boldsymbol\zeta/2)$ is not separable in
$\mathbf{r}_{bA}$ and $\boldsymbol\zeta$, so this step introduces
additional partial-wave couplings and requires a new numerical
module.
Several features make the computation genuinely difficult.
First, the six-dimensional integral over $\mathbf{r}_{bB}$ (3D) and
$\boldsymbol\zeta$ (3D) must be performed for each value of
$\mathbf{r}_x$; the non-separability prevents factorization into
independent radial integrals.
Second, the multipole expansion of $V_{bA} - U_{bA}$
(Sec.~\ref{sec:estimate}) converges slowly for a spatially extended
fragment such as the deuteron ($\zeta_{\mathrm{rms}} \approx 3.9$~fm,
comparable to the nuclear radius), so many partial waves are needed.
Third, the projectile wave function
$\varphi_a(\mathbf{r}_{bx}, \zeta)$ is itself a multi-coordinate
object: for reactions where $a$ is a three-body system
(e.g., ${}^6\mathrm{Li} = \alpha + d = \alpha + n + p$), computing
$\varphi_a$ already requires a three-body calculation.
A perturbative evaluation of the leading (dipole) correction may be
the most feasible first step, since the $E1$ matrix elements
$\langle\phi_0|\boldsymbol\zeta|\phi_n\rangle$ can be extracted from
CDCC bin states and the resulting source has a structure similar to
existing coupled-channel IAV calculations.
The full non-perturbative evaluation, including all multipoles, remains
an open computational challenge.

The fully microscopic problem is qualitatively harder still.
Once $V_{bA}$ carries explicit $\xi$-dependence and target excitations
are retained, the single-channel Green's function $G_x$ must be
replaced by the coupled-channel object $[\mathbf{G}_x]_{\alpha\beta}$,
and the calculation moves well beyond standard IAV infrastructure.
The analytical estimates of Sec.~\ref{sec:estimate} are therefore not a
substitute for a numerical calculation, but serve to identify the
physical regime where nonspectator effects matter and the multipoles
that dominate.

Within the reduced single-channel formulation, the decomposition into
spectator, cross, and nonspectator terms depends on the choice of the
reference potential $U_{bA}$.
The total source
$\tilde\rho_0^{\mathrm{tot}} = \tilde\rho_0 +
\tilde\rho_0^{(\mathrm{nsp})}$ and the total cross section are
independent of $U_{bA}$ (since the $U_{bA}$ terms cancel between
the two contributions), but the individual spectator and
nonspectator terms are not.
A natural choice is the adiabatic folding potential
$U_{bA} = U_{bA}^{\mathrm{fold}}$
[Eq.~(\ref{eq:Ufold})], which ensures that the diagonal correction
$\langle\phi_0|(V_{bA} - U_{bA})|\phi_0\rangle = 0$ vanishes and
the nonspectator source is driven entirely by off-diagonal
($\mathcal{Q} \to \mathcal{P}$) transitions.
If instead an empirical $b$-$A$ optical potential is used for
$U_{bA}$, one must be aware that such potentials are typically
fitted to elastic scattering data and may already absorb some
$\mathcal{P} \leftrightarrow \mathcal{Q}$ polarization effects
through their imaginary part and effective real
potential~\cite{Feshbach1992,Ichimura1985}.
In that case, the nonspectator correction $V_{bA} - U_{bA}$
should be understood as the residual beyond what the empirical
optical model already accounts for, and care must be taken to
avoid double-counting.

If $V_{bA} - U_{bA}$ is treated as a perturbation on the spectator
result, the NEB cross section can be expanded as
\begin{equation}
\sigma^{\mathrm{NEB}}
\approx
\sigma_{\mathrm{sp}}^{\mathrm{NEB}}
+ \sigma_{\mathrm{cross}}^{\mathrm{NEB},(1)}
+ \mathcal{O}\bigl[(V_{bA} - U_{bA})^2\bigr] \,,
\label{eq:perturbative}
\end{equation}
where $\sigma_{\mathrm{sp}}^{\mathrm{NEB}}$
[Eq.~(\ref{eq:sigma_sp_NEB})] is the composite spectator NEB cross
section and the first-order correction
$\sigma_{\mathrm{cross}}^{\mathrm{NEB},(1)}$ is the NEB part of the
cross term, obtained by applying the $G_x^\dagger W_x G_x$ projection
of Eq.~(\ref{eq:ImGx}) to the cross term
[Eq.~(\ref{eq:sigma_cross})].
For a given choice of $U_{bA}$, this provides a diagnostic of the
spectator approximation: if
$\sigma_{\mathrm{cross}}^{\mathrm{NEB},(1)}$ is small compared to
$\sigma_{\mathrm{sp}}^{\mathrm{NEB}}$, the spectator approximation
is justified for that reaction relative to the chosen reference
potential.

\section{Summary}
\label{sec:summary}

I have derived a generalization of the IAV inclusive breakup formalism
that removes the spectator approximation and provides state-resolved
inclusive cross sections for composite fragment $b$.
At the exact DWBA level, the generalized sum rule involves the full
$x + A$ resolvent $(E_{x,0}^+ - H_{xA})^{-1}$, while the familiar
single-channel IAV structure is recovered only after an additional
reduction that suppresses the explicit target dependence of $V_{bA}$.
Under a closure approximation, the standard IAV formalism with
structureless $b$ is most naturally associated with the total inclusive
cross section summed over all of $b$'s internal states, without
distinguishing whether the detected fragment is in its ground state
or has been excited or broken up.
The present formalism introduces the projection onto $b$'s ground
state explicitly, yielding the cross section $\sigma_0^{\mathrm{NEB}}$
for detecting $b$ intact, which is a distinct observable from the
total $\sigma_{\mathrm{total}}^{\mathrm{NEB}}$ to which the IAV
formalism corresponds under closure and structureless-reduction
approximations.
Within the DWBA, all nonspectator effects enter through the source
function via the operator $V_{bA} - U_{bA}$, while the propagator
is unchanged.
Because the source is modified before the EBU/NEB projection is taken,
nonspectator dynamics can in principle alter both the elastic and
nonelastic breakup components of the observed intact-$b$ signal.
Equivalently, the nonspectator correction is not synonymous with NEB;
it is a correction to the source that can feed either sector.
Post-prior equivalence is preserved.
A tidal estimate for $b = d$ on ${}^{208}\mathrm{Pb}$ shows that
the nonspectator operator is not a small perturbation at the
nuclear surface; whether this operator-level scale produces a
significant difference between $\sigma_0^{\mathrm{NEB}}$ and
$\sigma_{\mathrm{total}}^{\mathrm{NEB}}$ requires numerical
evaluation of the source integrals.

\begin{acknowledgments}
This work was supported by the National Natural Science Foundation
of China (Grant Nos.~12475132 and 12535009) and the Fundamental
Research Funds for the Central Universities.
Large language models were used to polish and condense portions of
the manuscript text and to check the consistency of the notation.
The author takes full responsibility for the scientific content
of this paper.
\end{acknowledgments}

\bibliography{references}

\end{document}